# Spectral approach to transport in the two-dimensional honeycomb lattice with substitutional disorder

E G Kostadinova [1], C D Liaw [1,2], A S Hering [3], A Cameron [4], F Guyton [5], L S Matthews [1], and T W Hyde [1]

[1] Center for Astrophysics, Space Physics & Engineering Research, Baylor University, Waco, TX, 76706, USA
[2] Department of Mathematical Sciences, University of Delaware, 311 Ewing Hall, Newark, DE 19716, USA
[3] Department of Statistical Science, One Bear Place 97310, Baylor University, Waco, TX, 76706, USA
[4] Department of Physics, Brigham Young University, Provo, UT 84602, USA
[5] Department of Physics, Applied Physics, and Astronomy, Rensselaer Polytechnic Institute, 110 8th St, Troy, NY 12180, USA

E-mail: Eva_Kostadinova@baylor.edu, liaw@udel.edu, Mandy_Hering@baylor.edu, adamc9900@gmail.com, guytof@rpi.edu, Lorin_Matthews@baylor.edu, Truell_Hyde@baylor.edu

**Abstract.** The transport properties of a disordered two-dimensional (2D) honeycomb lattice are examined numerically using the spectral approach to the quantum percolation problem, characterized by an Anderson-type Hamiltonian. In our simulations, substitutional disorder (or doping) is represented by a modified bimodal probability distribution of the on-site energies. The results indicate the existence of extended energy states for nonzero disorder and the emergence of a transition towards localized behavior for critical doping concentration $n_D > 0.3\%$, in agreement with the experimentally observed metal-to-insulator transition in graphene sheet doped with hydrogen.

## I. INTRODUCTION

Percolation theory is a simple, though not exactly solved, probabilistic model for studying phase transitions in disordered media. Mathematically, it describes the behavior of connected clusters in a random topological space. In physics, percolation theory was first introduced by Broadbent and Hammersley [1], who analyzed how the random properties of a porous medium influence the transport of a fluid moving through it. Since then, the percolation problem has often been used to model conductivity in disordered materials of various dimensions and geometry. With the discovery of graphene and related two-dimensional crystals [2], [3], the study of transport in the 2D honeycomb lattice attracted considerable attention [4], [5].

According to the well-established scaling theory [6], any nonzero amount of disorder in a 2D crystal (characterized by an Anderson-type Hamiltonian) leads to either exponential or logarithmic localization of the wave function propagating through the medium. In other words, there is no real metal-to-insulator transition (MIT) in a 2D disordered lattice. Similarly, it has been postulated that a wavefunction cannot percolate through a two-dimensional system unless all energy sites in the corresponding graph are open with probability 1, which is analogous to a lattice with zero disorder [1]. Nevertheless, MIT has been experimentally observed for graphene doped with $NO_2$ [7], hydrogen [8], and boron [9]. It has been argued [10]–[12] that the "apparent" existence of extended states in doped graphene results from the material's long localization length, which often extends beyond the size of the examined sample. An alternative explanation is that finite-size scaling methods, which are often adopted in the analysis of 2D transport problems, impose restrictions on the Hilbert space that exclude *a priori* the existence of extended states. Thus, there is a need for a mathematical treatment of the problem where the entire Hilbert space is considered.

In this paper, we employ recent results from spectral theory in the analysis of quantum percolation on the 2D doped honeycomb lattice. The spectral approach (first introduced by Liaw in [13]) is a novel mathematical technique that can determine (with probability 1) the existence of extended energy states in infinite disordered lattices characterized by Anderson-type Hamiltonians. In this method, the spectrum of a bounded self-adjoint Hamiltonian (acting on the separable Hilbert space) is decomposed into absolutely continuous (ac) and singular parts. The presence of an absolutely continuous part corresponds to delocalization of the wavefunction propagating through the medium. Unlike techniques based on finite-size scaling, the spectral approach does not require restrictions on the Hilbert space, periodic boundary conditions, or the use of perturbation theory. Thus, it is ideal for the analysis of Anderson-type and percolation problems, allowing insight into the critical 2D case, where theoretical predictions and experimental results often disagree.

The spectral approach has been previously used to show that delocalization can exist in 2D square, triangular, and honeycomb lattices, when the random on-site energies (representing disorder) are selected from a rectangular (or continuous uniform) probability distribution [13]–[15]. This choice corresponds to the classical zero-temperature Anderson localization problem (first introduced in [16]). In the present study, on-site random energies are assigned according to a modified bimodal probability distribution designed to model the quantum percolation problem on 2D honeycomb lattice. Our results indicate that transport in the form of extended energy states can exist for small doping concentrations. We further confirm that the onset of metal-to-insulator transitions occur for doping levels of $\approx 0.3\%$, as experimentally observed in [8].

Section II, we briefly introduce quantum percolation as a limiting case of the binary alloy model of doping. The theoretical background provides motivation for our choice of a probability distribution for the random variables in the 2D crystal. In Section III, we describe how the spectral approach is applied to the discrete random Schrödinger operator with on-site energies selected from the modified bimodal distribution. The numerically predicted transition region is then compared against results from experiments employing doped graphene (Section IV). Finally, in Section V, we summarize the conclusions of this study and outline future research directions.

## II. THEORETICAL BACKGROUND

---

[1] The connection between Anderson localization and quantum percolation will be explored in more detail in Sec II A 3.

## A. Formulation of the transport problem

In this study, we are interested in the critical amount of substitutional disorder (or doping) sufficient to induce a phase transition in the transport properties of the 2D honeycomb crystal. In the tight-binding approximation, the single-electron, noninteracting Hamiltonian on the 2D honeycomb lattice $\Lambda$ has the form

$$H = \sum_{\substack{i,j \in \Lambda \\ i \neq j}} |e_i\rangle V_{ij} \langle e_j| + \sum_{i \in \Lambda} |e_i\rangle \epsilon_i \langle e_i|, \tag{1}$$

where $|e_i\rangle$ are any set of basis vectors of the 2D space $\Lambda$, and $V_{ij}$ is the hopping potential [2] between pairs of nearest neighbors. The on-site energies $\epsilon = \{\epsilon_i\}_{i \in \Lambda}$ form a set of independent variables chosen from an interval $[a, b]$ according to a prescribed probability density distribution $\chi(\epsilon)$. The probability that $\epsilon_i$ is selected from any subinterval $[a', b'] \in [a, b]$ is given by the area under the curve between the points $a'$ and $b'$, that is,

$$A = \int_{a'}^{b'} \chi(\epsilon) d\epsilon. \tag{2}$$

Assume the on-site energies in the unperturbed crystal are constant (say, zero) and that the hopping potential is a constant (here $V_{ij} = V = 1$). Then, assigning the variables $\{\epsilon_i\}$ to the lattice sites corresponds to introducing impurities into the crystal. If the area $A$ defined in (2) is normalized to unity, then every lattice site is assigned an energy value from the prescribed $\chi(\epsilon)$. In this formulation, the amount and type of disorder can be varied by changing the width of the interval $[a, b]$ and by altering the shape of $\chi(\epsilon)$, respectively.

In our previous work [14], [17], we examined the standard Anderson localization problem, where the variables $\{\epsilon_i\}$ are independent and identically distributed (i.i.d.) in the interval of allowed energies according to a rectangular probability distribution

$$\chi(\epsilon) = \begin{cases} 0, & \epsilon \notin [a, b] \\ 1/W, & \epsilon \in [a, b] \end{cases}, \tag{3}$$

where $W$ is the width of the interval $[a, b]$. When distribution (3) is used, each energy value $\epsilon \in [a, b]$ is occurs with equal probability, and the amount of disorder is controlled by the width $W$ of the interval $[a, b]$. Thus, an important feature of this type of defect (which we call an Anderson-type defect) is randomness that can be physically realized as random spacing of impurities or random arrangement of electronic / nuclear spins in the examined system [16].

If the on-site energies of the unperturbed crystal are not constant but exhibit slight variations (due to thermal fluctuations or stochastic effects), the probability distribution of possible energy states in a single-species crystal can be modeled using a Gaussian function the tails of which represent the deviation from the mean energy characteristic of that species (Fig. 1a). In this case, a variation of the disorder can be achieved by changing both the skew of the distribution (for example, through multiplication by exponential function) and the width of the peak. In the limit of an ideal lattice at zero temperature, the Gaussian distribution should approach a delta function [3].

---

[2] The hopping potential is the kinetic energy term allowing for electron tunneling or *hopping* between nearest neighbor lattice sites.

In the presence of a second atomic species, it is convenient to use the bimodal distribution (Fig. 1b), which consists of two continuously connected Gaussian peaks. This probability distribution corresponds to a mixture of two species, where the $A$-type atoms have a characteristic energy $E_A$ (the mean value of the first peak) and $B$-type atoms have characteristic energy $E_B$ (the mean value of the second peak). In the limit where $E_B - E_A \to \infty$, one species acts as an open state, while the other species acts like a perfect barrier. This scenario is often modeled by the classical percolation problem. If the difference $E_B - E_A$ is large but finite, the bimodal distribution can be used to study the quantum percolation problem, which belongs to the same universality class as does the Anderson localization problem.

In the present numerical analysis, doping is assigned according to a modified bimodal distribution, which consists of a single Gaussian peak centered at energy $E_A = 0$ and a delta function peak centered at energy $E_B = 100$. The motivation for this choice of probability distribution and its connection to the quantum percolation and Anderson localization problems are examined in the following sections.

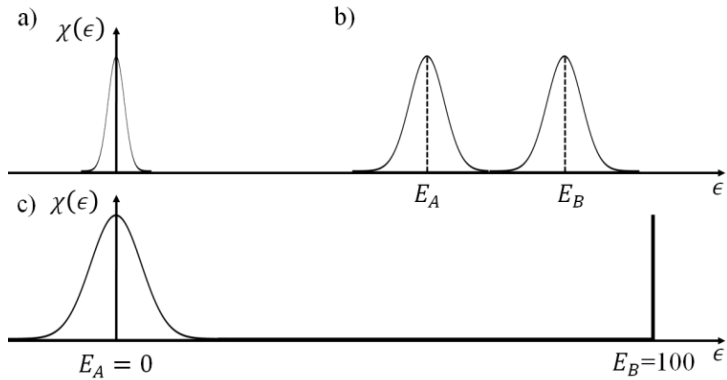

FIG. 1. Graphs of the a) Gaussian, b) bimodal, and c) modified bimodal probability density distributions as discussed in the text. Note that the distribution in c) is obtained from the one at b) by keeping the width of one Gaussian fixed and shrinking the width of the other so that the latter approaches a delta function. Thus, although the two peaks have large separation, they still form a continuous distribution function.

*1. Binary alloy model of doping*

Consider a 2D crystal composed of $A$-type atoms with (most probable) energy $E_A$ and $B$-type atoms with (most probable) energy $E_B$, where $E_A \neq E_B$. In general, such a system can be modeled by the binary alloy distribution

$$\chi(\epsilon_i) = p\delta(\epsilon_i - E_A) + (1-p)\delta(\epsilon_i - E_B), \tag{4}$$

where $\delta(\epsilon_i - E_A)$ and $\delta(\epsilon_i - E_B)$ are the unimodal probability distributions of the $A$-atoms and the $B$-atoms, respectively, and $p$ is the mixing parameter. Here we consider a mixture of two Gaussian peaks (see Fig. 1(b))

$$\delta(\epsilon_i - E_A) = \frac{1}{\sqrt{2\pi\sigma_A^2}} exp\left[-\left(\frac{\epsilon_i - E_A}{2\sigma_A^2}\right)^2\right] \tag{5a}$$

and

---
[3] Note that due to the Pauli exclusion principle, the possible energies in a crystal must form a band, i.e., the width of the Gaussian function can never reach zero.

$$\delta(\epsilon_i - E_B) = \frac{1}{\sqrt{2\pi\sigma_B^2}} exp\left[-\left(\frac{\epsilon_i - E_B}{2\sigma_B^2}\right)^2\right]. \quad (5b)$$

In this case, the mixing parameter $p$ corresponds to the concentration (fraction) of $A$-atoms, while $(1-p)$ is the concentration of $B$-atoms. For appropriate choices of mixing parameter $p$ and characteristic energies $E_A$ and $E_B$, the binary alloy formulation represents a doped crystal with substitutional disorder. Assuming fixed values for the means $E_A$ and $E_B$, one can vary the standard deviations $\sigma_A$ and $\sigma_B$ of each Gaussian peak and / or the mixing parameter (concentration) $p$ until critical behavior is observed.

*2. Quantum percolation problem*

Consider the 2D integer lattice, where each pair of nearest neighbor sites (called vertices) is connected by a bond. In the site percolation problem [4], all bonds are considered open, while the lattice sites are, independently of each other, chosen to be open with probability $p$ and closed with probability $(1-p)$. Denote $C(0)$ as the open cluster (set of open vertices) that contains the origin. The percolation function $\theta(p)$ is the probability that there exists an open cluster $C(0)$ that extends from the origin to infinity, i.e., the probability that the system percolates. Since $\theta(p=0) = 0$ (all vertices closed) and $\theta(p=1) = 1$ (all vertices open), there exists a critical value $p_c$ for which the transport behavior of the lattice exhibits a phase transition.

The probability distribution for the quantum site-percolation problem can be obtained from equation (4) in the limit $E_B \to \infty$ [18], [19], which gives a single Gaussian distribution function

$$\chi(\epsilon_i) = p\delta(\epsilon_i) = p\frac{1}{\sqrt{2\pi\sigma_A^2}} exp\left[-\left(\frac{\epsilon_i - E_A}{2\sigma_A^2}\right)^2\right]. \quad (6)$$

Without loss of generality, one can further choose $E_A = 0$, which centers the peak at the origin. In this formulation of the problem (known as the classical site percolation), the $A$-type atoms are open (or perfect acceptors), while the $B$-type atoms are closed (perfect barriers). Thus, transport occurs only on a random assembly of $A$-atoms, and the existence of extended or localized clusters is dependent on variation of the concentration $p$.

However, in a quantum mechanical system, there is a finite probability for tunneling to each lattice site, and one cannot assume the existence of perfect barriers. Instead, the characteristic energy $E_B$ is represented by a large (yet finite) number. It is known [20] that when $E_B - E_A > 2ZV$ (where $Z$ is the number of nearest neighbors and $V$ is the hopping potential), the spectrum of the Hamiltonian given in Eq. (1) splits into two sub-bands centered approximately around $E_A$ and $E_B$. For the 2D honeycomb lattice used in our simulations, we let $Z = 3$ and $V = 1$. Thus, a choice of $E_B - E_A = 100$ ensures that the two bands are separate and far apart from one another, yet a finite distance away. In other words, lattice sites with energy $E_B$ are unfavorable to occupy but do not represent forbidden states.

*3. Transport in the critical 2D case*

Let the critical probability in the percolation problem be denoted $p_c$ in the classical regime and $p_q$ in the quantum regime. Unlike the classical percolation problem, the quantum case accounts for the effect of quantum interference. As the particle passes along various routes in the disordered medium, it accumulates different phases. Destructive interference of such phases can halt the diffusion of the particle's wave function, which, in principle, is also the mechanism causing localization in the Anderson problem. Since it has been shown that quantum percolation and Anderson localization belong to the same universality class of transport phenomena [21], [22], it is straightforward to outline a few parallels in the logic of the two models.

---
[4] Alternatively, one can consider all vertices to be open and let the bonds be open or closed with a certain probability. This setup is called a *bond percolation* problem.

In the subcritical phase of the quantum percolation problem (i.e., $p \leq p_q$), it is known that $\theta(p) = 0$ and $C(0)$ is finite with probability 1. The corresponding transport behavior in the Anderson model has localization length that is exponentially proportional to the reciprocal of the disorder strength. Similarly, in the supercritical phase (i.e., $p > p_q$), $\theta(p) > 0$ and there is a finite probability that $C(0)$ spreads to infinity, which corresponds to delocalization of the particle wave function. Around the critical point $p = p_c$, numerous functions of interest exhibit power law behavior, which is also a characteristic of functions approaching the critical amount of disorder $W_c$ where the metal-to-insulator transition occurs in the Anderson model. Although there is no trivial correspondence between $p_q$ and $W_c$, it is expected that one is a function of the other (due to the equivalency of the two transport problems). Thus, it is not surprising that the question of whether a 2D system can percolate for $p_q < 1$ has been a decades-long debate closely related to the disagreement on the existence of extended states for nonzero disorder in the 2D Anderson problem.

Earlier numerical studies [23]–[25] based on finite-size scaling predicted that in the 2D case, the system can percolate only if all states are open, i.e., at $p_q = 1$. Later it was suggested [19] that percolation occurs for $p_q < 1$ but at energies far away from the band center ($E = 0$). Proponents of these results were largely encouraged by the remarkable success of the scaling approach to the Anderson localization problem [6]. Another fraction of the physicists and mathematicians argued that the 2D lattice can percolate for $p_q < 1$ at all energies [26]–[29]. However, the exact value of the critical probability $p_q$ appears to be sensitive to the applied technique. To illustrate this point, Table I presents a selection of $p_q$ values obtained for the square lattice, which is the most studied two-dimensional geometry. For the less researched honeycomb case examined here, it was suggested that the value of the percolation threshold in the quantum regime varies with system size [4], leaving the question of the critical point unsettled.

Table I. Critical probability value $p_q$ in the 2D quantum site percolation problem on the square lattice.

| Authors | Method | $p_q$ |
|---|---|---|
| Odagaki et al. [30] | Green's function method | 0.59 |
| Koslowski & von Niessen [31] | Thouless-Edwards-Licciardello method | 0.70 |
| Srivastava & Chaturvedi [29] | Method of equations of motion | 0.73 |
| Daboul et al. [28] | Series expansion methods | 0.74 |
| Odagaki & Chang [32] | Real-space renormalization group method | 0.87 |
| Raghavan [33] | Mapping a 2D system into a one-dimensional system | 0.95 |

For a 2D lattice of any geometry, it has been shown [19], [22], [23], [34] that the graph of $\theta(p)$ is flat in the subcritical phase and at the critical point, i.e. for $p \leq p_q$. It is also expected that the phase transition from a finite to infinite cluster size in 2D has no jump discontinuities, although the shape of the graph of $\theta(p)$ in the region $p_q < p < 1$ is still conjectural. If scaling theory arguments are valid, then $p_q = 1$ and the graph of $\theta(p)$ is a flat line with a jump discontinuity at the critical point (as shown by the blue line in Fig. 2). In contrast, if $p_q < 1$, the transition region is assumed to be continuous, but the shape of the graph in the supercritical regime needs further investigation. As mentioned above, numerical studies yielding $p_q < 1$ often do not provide rigorous predictions regarding the shape of the graph in the transition region, making them hard to apply to physical systems. We further note that the existence of $p_q < 1$ does not imply transport with probability 1; rather it suggests that in the interval $p_q < p < 1$, the function $\theta(p)$ is increasing (probably) continuously from $\theta(p_q) = 0$ to $\theta(p) = 1$. Thus, the shape of the percolation function of the 2D system in the supercritical regime cannot be established with certainty.

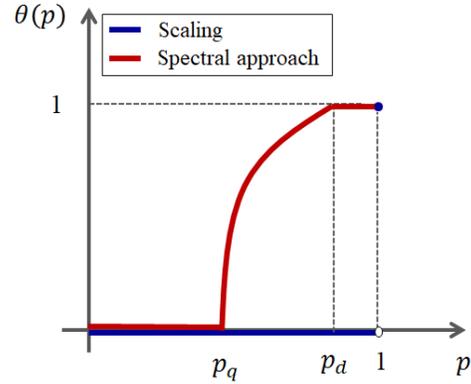

The application of the spectral approach to the 2D quantum percolation problem provides an important improvement on this issue. In our previous work [13], [14], [17], we established the existence of transport in 2D lattices having nonzero Anderson-type disorder. Here, we similarly confirm that a 2D quantum system percolates for $p_q < 1$. Unlike most of the methods mentioned above, the spectral approach shows the existence of percolation with probability 1. In other words, it identifies the critical region for values of $p$ where $\theta(p) = 1$ (and the graph becomes a flat line). We define the lower bound of this region by $p_d$, where $d$ stands for delocalization. Note that $p_q$ and $p_d$ do not have to coincide as illustrated by the red line in Fig. 2.

FIG. 2. $\theta(p)$ vs. $p$ according to scaling theory (blue line) and according to the spectral approach (red line). Note that below the critical point $p_q$, the function $\theta(p) = 0$, while above the critical point $p_d$, the function $\theta(p) = 1$. The shape of the graph in the transition region $p_q < p < p_d$ is conjectural.

### B. Spectral approach

We now provide a brief discussion of spectral theory and its application to transport problems. The basic steps in the spectral method are introduced for the specific case of a disordered 2D honeycomb lattice $\Lambda$ (Fig. 3), characterized by the discrete random Schrödinger operator. It is important to note, however, that the spectral approach can be applied to the entire class of Anderson-type Hamiltonians (first introduced in [35]) and generalized to any dimension or geometry. Definitions of mathematical concepts, such as cyclicity and spectral decomposition, can be found in [13], [14], [36]. These references also provide detailed proofs and a physical interpretation of the model.

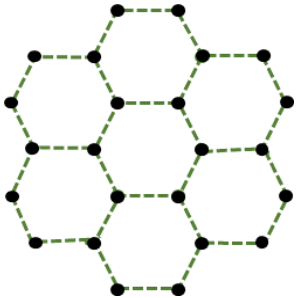

FIG. 3. Honeycomb lattice $\Lambda$, where black dots represent lattice vertices and dashed lines represent bonds.

#### 1. Spectral decomposition of the Hamiltonian

The 2D honeycomb lattice $\Lambda$ is the graph $G = (V, E)$ of the 2D set of vertices (or sites) $\{v_i\}$ connected by edges (or bonds) $\{e_{ij}\}$. Bonds represent the graph distance between pairs of nearest neighbors (i.e., vertices located at a Euclidean distance 1 apart from each other). For any function of the vertices $f(v_i)$ taking values in the 2D set of all integers $\Lambda$, the (discrete) Laplacian acting on $f$ is given by

$$\Delta f = \sum_{e_{ij}=1} [f(v_i) - f(v_j)], \qquad (7)$$

where $e_{ij}$ is the edge between vertices $v_i$ and $v_j$, and the sum is over the nearest neighbors of vertex $v_i$. Thus, we see that the discrete Laplacian $\Delta$ is the graph representation of the hopping potential term in equation (1).

Let $l^2(\Lambda)$ be the Hilbert space of square-summable sequences on the 2D space $\Lambda$. In this case, the graph representation of the Hamiltonian given in equation (1) is the discrete random Schrödinger operator on $l^2(\Lambda)$

$$H_\epsilon = -\Delta + \sum_{i \in \Lambda} |e_i\rangle \epsilon_i \langle e_j|, \qquad (8)$$

where $\epsilon_i \in [a, b]$ are defined as in Sec. II A and the set of base vectors $|e_i\rangle$ now represent the vertices of the discrete lattice $\Lambda$. Without loss of generality, assume that the interval of possible on-site energies has a width $W$ and is symmetric about the origin, i.e., $\epsilon_i \in [-W/2, W/2]$. Thus, the value of the parameter $W$ can be used to vary the amount of disorder in the crystal. The random Schrödinger operator given in (8) models a honeycomb lattice comprised of atoms located at the vertices.

The spectral approach uses the Spectral Theorem to diagonalize the Hamiltonian and obtain its energy spectrum. The Spectral Theorem maps the square summable Hilbert space $l^2(\Lambda)$ to a new space $L^2(\mu)$, consisting of the square-integrable functions with respect to the spectral measure $\mu$. This new space $L^2(\mu)$ can be decomposed into two orthogonal Hilbert spaces, $L^2(\mu_{ac})$ and $L^2(\mu_{sing})$, where the spectral measure $\mu$ is decomposed into an absolutely continuous part and a singular part

$$\mu = \mu_{ac} + \mu_{sing}. \qquad (9)$$

The energy spectrum of the Hamiltonian therefore has a singular part $(H_\epsilon)_{ac}$ that comes from $L^2(\mu_{ac})$ (representing the extended states) and a part $(H_\epsilon)_{sing}$ that comes from $L^2(\mu_{sing})$ ( roughly speaking, corresponding to localized states). It was shown [35] that cyclicity of vectors in $l^2(\Lambda)$ is related to the singular part of the spectrum and that non-cyclicity of any vector in the same space indicates the existence of an absolutely continuous part of the spectrum.

*Theorem* [37]: For any *nonzero* vector $v_0$ [5] in the lattice space $l^2(\Lambda)$, $v_0$ is cyclic for the singular part $(H_\epsilon)_{sing}$ with probability 1.

*Theorem* [13]: If one shows that $v_0$ is *not* cyclic for $H_\epsilon$ with non-zero probability, then almost surely [6]

---

[5] Note that, in this section, a vertex $v_i$ is the graph analogue of a base vector $|e_i\rangle$ while $v_i$ is any vector in the space (i.e. $v_i$ is a linear combination of the chosen base vectors).
[6] Note that in probability theory an event happens *almost surely* if it happens *with probability* 1. In this paper, we use the two phrases interchangeably.

$$H_\epsilon \neq (H_\epsilon)_{sing}, \tag{10}$$

which indicates the existence of extended states since $H_\epsilon$ must have a non-zero absolutely continuous part.

*2. Spectral approach*

For a given realization of the disorder in the 2D honeycomb lattice:

(i) Fix a random vector, say $v_0$, in the 2D space $l^2(\Lambda)$, and generate the sequence $\{v_0, H_\epsilon v_0, (H_\epsilon)^2 v_0, \cdots, (H_\epsilon)^N v_0\}$ where $N \in \{0,1,2,\ldots\}$ is the number of iterations of $H_\epsilon$ and is used as a timestep.

(ii) Apply a Gram-Schmidt orthogonalization process (without normalization) to the members of the sequence, and denote the new sequence $\{m_0, m_1, m_2, \ldots, m_N\}$.

(iii) Calculate the distance from *another* vector in the lattice space, say $v_1$, to the $N$-dimensional orthogonal subspace $\{m_0, m_1, m_2, \ldots, m_N\}$, given by [13]

$$D_{\epsilon,W}^N = \sqrt{1 - \sum_{k=0}^{N} \frac{\langle m_k | v_1 \rangle^2}{\|m_k\|_2^2}}, \tag{11}$$

where $\|\cdot\|_2$ is the Euclidean norm and $\langle \cdot | \cdot \rangle$ is the inner product in the space. It can be shown [13], [36] that if for any $v_1$ (with nonzero probability) the limiting behavior of the distance vector obeys

$$\lim_{N \to \infty} D_{\epsilon,W}^N > 0, \tag{12}$$

then extended states exist with probability 1.

### III. NUMERICAL WORK

The spectral method of Sec. II B was applied to the discrete random Schrödinger operator given in equation (8). In all simulations presented here, we assume $\Delta = ZV$ (i.e., constant hopping potential over the nearest neighbors), and let $\{e_i\}_{i \in \Lambda}$ be the standard basis vectors of the honeycomb lattice, such that $e_i$ assumes the value 1 in the $i^{th}$ entry and zero in all other entries [7]. With this assumption, the Hamiltonian becomes

$$H_\epsilon = -ZV + \sum_{i \in \Lambda} \epsilon_i e_i \langle e_i|. \tag{13}$$

The numerical analysis starts by generating one realization of the random variables $\epsilon_i$ according to the prescribed modified bimodal distribution $\chi(\epsilon)$. Next, a random base vector $e_0$ is selected and used to produce the sequence $\{e_0, He_0, H^2 e_0, \ldots, H^N e_0\}$, where $N = 4500$ corresponds to the number of iterations of the Hamiltonian and is used as a timestep. Then the orthogonalized sequence $\{m_0, m_1, m_2, \ldots, m_N\}$ is generated using the Gram-Schmidt procedure (without normalization). At each timestep, the corresponding distance value $D_{\epsilon,W}^N$ is obtained using equation (11). Finally, the graph of $D_{\epsilon,W}^N$ vs. $N$ is analyzed to determine the limiting behavior of $D_{\epsilon,W}^N$ as $N \to \infty$.

---

[7] Note that the spectral approach only requires that $v_0$ and $v_1$ be any two (different) vectors in the Hilbert space of interest. The choice $v_0 = e_0$ and $v_1 = e_1$ ensures faster computation times and is not related to the generality of the method.

To ensure accuracy of the numerical results, we perform an orthogonality check on the sequence $\{m_0, m_1, m_2, ..., m_N\}$, which indicates that the obtained vectors are indeed orthogonal, i.e., there is no Gram-Schmidt instability in the algorithm. A discussion on the orthogonality check procedure can be found in [17].

The use of the modified bimodal distribution for the random variables $\{\epsilon_i\}$ allows for introduction of two distinct types of defects: positional and substitutional. The positional disorder is controlled by the variance $\sigma^2$ of the Gaussian peak [8], while doping is achieved by variation of the concentration [9] $n_D$ of $B$-type lattice sites of energy $E_B$. In Sec. III A, we explain the suitability of the distribution variables $E_A$, $E_B$, and $\sigma^2$ for the present study by examining their effect on the outcomes of the spectral method. In Sec. III B, we fix these parameters and explore the influence of increasing the doping $n_D$ on the transport behavior of the 2D honeycomb lattice.

### A. Distribution variables

In each simulation, the modified bimodal distribution is characterized by one Gaussian peak centered at $E_A = 0$ and a second peak which is (approximately) a delta function located at $E_B = 100$. Here the delta function represents dopant atoms with substantially higher average energy $E_B$. Since the introduction of substitutional disorder in a material is usually a controlled process, we do not consider a spread of possible energies for the $B$-type atoms. In other words, the defect is produced by a careful substitution of an $A$-type atom with a $B$-type atom, whose energy is controlled.

In Sec. II A 2, we argued that the chosen energy difference $E_B - E_A = 100$ is sufficient to represent a quantum percolation problem, where the $B$-type atoms are unfavorable but not forbidden lattice sites. To ensure that the choice $E_A = 0$ can be used without loss of generality, we applied the spectral approach to a 2D honeycomb lattice where the random energies were assigned according to a normal distribution (single Gaussian peak) with fixed variance $\sigma^2 = 0.4$ and mean energy varied within the ranges $E_A \in [0,100]$ and $E_A \in [0,1]$. Representative distance plots corresponding to values from the two ranges are shown in Fig. 4a and Fig. 4b. Comparison of the plots indicates that for a fixed variance, the limiting behavior of $D_{hc}$ [10] is not altered appreciably as the mean energy of the normal distribution is varied. Specifically, even for the two extremes of the examined intervals ($E_A = 0$ and $E_A = 100$), Fig. 4a shows that the behavior of the corresponding distance parameters follows a very similar trend. This confirms that the spectral analysis is not affected by the choice of mean energy for the normal distribution. Thus, in the following simulations, the Gaussian peak in the modified bimodal distribution is centered at $E_A = 0$, which is in agreement with previous numerical simulations [18], [19], where the normal distribution was used to model quantum percolation. The choice $E_A = 0$ also makes sense physically since it represents the most probable (or expected) value, which, in the unperturbed crystal, should correspond to a minimum in the energy band.

---

[8] In the following sections, we let $\sigma_A^2 \rightarrow \sigma^2$ for notational simplicity.
[9] Here $n_D$ stands for the concentration of the $B$-type doping material and is equal to the probability for a closed barrier $1 - p$ in the quantum percolation problem.
[10] The subscript $hc$ stands for *honeycomb*.

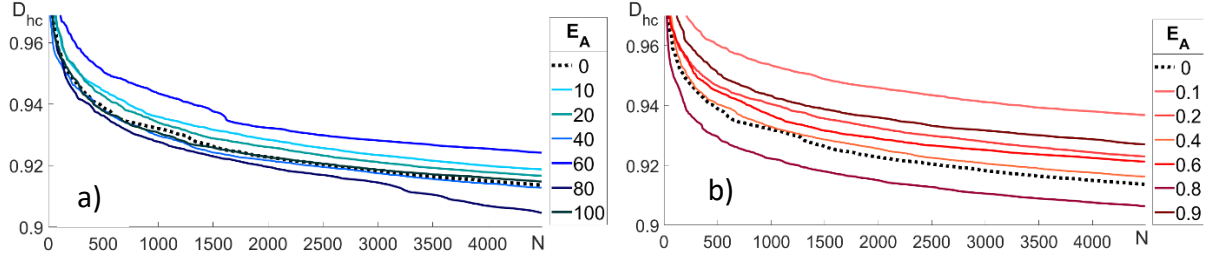

FIG. 4. Representative $D$-plots generated assuming a Gaussian distribution with fixed variance $\sigma^2 = 0.04$ and mean varying in the ranges a) $E_A \in [0,100]$ and b) $E_A \in [0,1]$. In both cases, as the mean is increased from zero, the limiting behavior of $D_{hc}$ does not deviate significantly. The case $E_A = 0$ is shown as a black dotted line on the graph.

To evaluate how changing the variance $\sigma^2$ influences the transport behavior of a system, we applied the spectral method to lattices generated using a Gaussian distribution with mean $E_A = 0$ and variance within the range $\sigma^2 \in [0.05, 0.50]$. The distance plots in Fig. 5a show that as the variance is increased, the slopes become increasingly negative. Since delocalization is established when $\lim_{n \to \infty} D_{hc} \neq 0$ (see Sec. II B 2), the existence of extended states is more likely for plots which quickly approach a zero slope. In contrast, as the slope of the $D_{hc}$-plots becomes more negative, the likelihood of $D_{hc}(\infty) > 0$ decreases. To quantify this reasoning, we require that delocalization can be claimed for plots that flatten exponentially for the given numerical timestep ($N = 4500$). In Fig. 5b, we see that for $\sigma^2 \leq 0.15$, the log-log $D_{hc}$-plots are straight lines with almost no slope, which indicates exponential decay, while for $\sigma^2 > 0.15$, the negative slope of the lines monotonically increases with increasing $\sigma^2$.

The determination of an exact transition point in the transport behavior of a lattice with on-site energies assigned from a Gaussian distribution is not the focus of this work. Here, we simply note that for a Gaussian distribution with a fixed mean $E_A = 0$, changes in the variance significantly affect the behavior of the distance parameter. This is to be expected since, physically, the variance quantifies the deviation from the mean energy and can be interpreted as a type of disorder in the system. Note, however, that in this case, the spread of the Gaussian function represents a defect characteristic of the undoped crystal and is, thus, similar to the Anderson-type disorder.

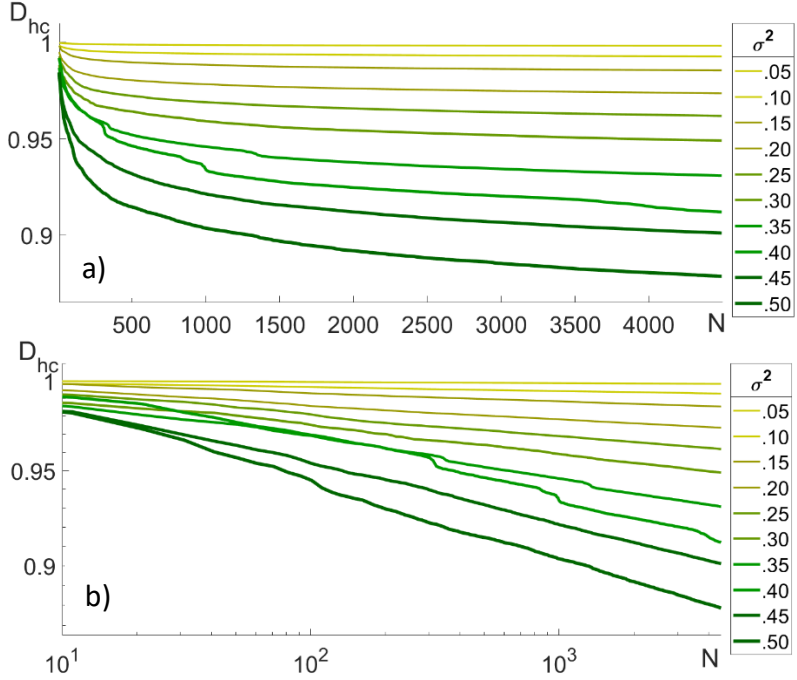

FIG. 5. Distance time evolution plots a) and corresponding log-log plots b) for the Hamiltonian in equation (13), where the random variables are assigned according to a Gaussian distribution with fixed mean $E_A = 0$ and variance in the range $\sigma^2 \in [0.05, 0.50]$.

For each value of $\sigma^2$, $D_{hc}$ is an average of five realizations of the on-site energies, which minimizes computational errors.

Although the focus of this work is substitutional defects, a realistic model of disorder should also account for lattice imperfections in the unperturbed crystal. Thus, it is useful to assume a value for $\sigma^2$ that contributes to the total effect of impurities but does not dominate it. Since the normal distribution models a type of defect, which is similar to the Anderson-type disorder, we obtain an estimate for the appropriate value of $\sigma^2$ through comparison with our previous study [17], where we established that for the 2D honeycomb lattice with Anderson-type disorder (i.e., rectangular distribution), extended states exist for $W \leq 0.75$. In this study, it is important to choose a variance that corresponds to $W$ smaller than this critical value.

An approximate relationship between the width of the square distribution $W$ and the variance of the normal distribution $\sigma^2$ is presented in Fig. 6. Recall that $\approx 95\%$ of the area of the normal distribution falls within a distance $2\sigma$ from the mean, so it is standard to approximate the Gaussian by a square distribution with width $W = 4\sigma$ (light pink square in Fig. 6) or $W^2 = 16\sigma^2$. Therefore, the critical value $W \leq 0.75$ approximately corresponds to $\sigma^2 \leq 0.035$. The same calculation performed with a square of width $2\sigma$ (light blue square in Fig. 6) yields $\sigma^2 \leq 0.14$. Clearly, the pink square overestimates and the blue square underestimates the actual area of the Gaussian. Thus, it is appropriate to choose the variance from the range $0.035 < \sigma^2 < 0.14$.

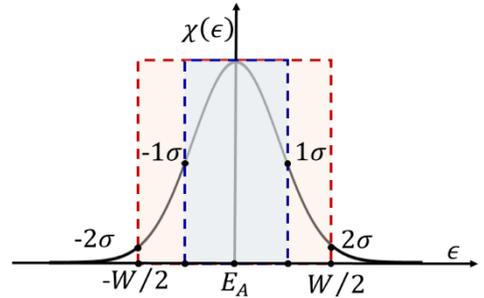

FIG. 6. Visual representation of the connection between $W$ for a square distribution and $\sigma$ for a Gaussian distribution.

Based on the above, we chose $E_A = 0$ and $\sigma^2 = 0.05$ for the Gaussian peak of the modified bimodal distribution.

B. 2D honeycomb lattice with substitutional disorder

We now consider the modified bimodal distribution with a Gaussian peak (mean $E_A = 0$ and variance $\sigma^2 = 0.05$) and a delta function at $E_B = 100$. Here, $n_G = p$ is defined as the concentration of lattice sites with energy $\epsilon_i$ selected from the Gaussian peak and $n_D = 1 - p$ is defined as the concentration of lattice sites with energy $E_B = 100$. In each simulation, we assign on-site energies from the normal distribution to obtain a single-species crystal with small lattice imperfections. We then simulate substitutional disorder by randomly assigning the energy $E_B$ to a fraction of the lattice sites determined by the value $n_D$. The doping concentration was varied over two ranges, corresponding to large doping, $n_D(\%) \in [1,32]$ and small doping $n_D(\%) \in [0,1)$. To reduce the effect of numerical errors, for each value of $n_D$, we generated multiple realizations (30 for large and 50 for small doping) of the on-site energies and then averaged the corresponding distance values.

*1. Large variation of doping concentration*

The time evolution of $D_{hc}$ for $n_D(\%) = 1:1:10$, plotted in Fig. 7a, indicates that for doping as small as 2% the distance values show an observable drop from the undoped lattice case $n_D(\%) = 0$ (shown as a black dashed line on the graph). Similarly, the corresponding log-log plots exhibit increasingly negative slopes for $n_D(\%) \geq 0.2$, which suggests that these $D_{hc}$ values do not flatten exponentially over the number of iterations in the present simulation. In addition, as the doping concentration increases, some realizations rapidly drop to zero, which results in sharp drops in the averaged distance plots (more visible as kinks in Fig. 7b). According to the spectral approach, the occurrence of such a realization (which we define as 'failures') indicates that delocalization cannot be established for the corresponding disordered system. Thus, the method suggests that for doping $n_D(\%) \geq 0.2$, the 2D honeycomb lattice has already transitioned into an insulating behavior, which agrees with experimental results. The consistency of these predictions is further supported by data shown in Fig. 7c, where the number of failed realizations is plotted as a function of increasing doping concentration in the range $11 \leq n_D(\%) \leq 32$.

Note that, even for the cases where all realizations for a given $n_D$ have a positive value at the last timestep of the simulation, i.e., $D_{hc}(4500) > 0$, proof of existence of extended states requires examination of the limiting behavior as $N \to \infty$. To extrapolate $D_{hc}(N \to \infty)$, we fit the data using the equation

$$D_{hc} = mN^{-\alpha} + b, \tag{14}$$

where the first term shows how rapidly $D_{hc}$ approaches a constant value $b$ as $N \to \infty$. Since all plots exhibit small fluctuations over the first thousand timesteps, we apply a nonlinear fitting to equation (14) using a weight function of the form $w = 1/\sqrt{4500 - n}$. In this way, the fit to the initial data reduces the effect of fluctuations, and the more accurately reflects the behavior at large $N$. To evaluate the contribution of the first term in equation (14), we define the ratio parameter

$$R = \frac{D_{hc}(4500) - b}{D_{hc}(4500)} \times 100\%, \tag{15}$$

which gives the percent contribution of the decaying term at the last timestep of the simulation.

Mathematically, showing that $D_{hc}$ limits to *any* nonzero positive number (however small) is sufficient evidence for the existence of extended states with probability 1. However, due to the finite character of the numerical simulations (restricted by the computation time), it is necessary to introduce a criterion for the maximum acceptable value for the ratio $R$. A rough estimation can be performed using the following physical reasoning. The chosen number of iterations, $N = 4500$, yields a lattice size of $\approx 4 \times 10^7$, which (using the average carbon-carbon bond length for graphene, $a \approx 0.142$nm) corresponds to a sample of the 2D material with area $\approx 1\mu m^2$. The sizes of graphene flakes (currently available for industry) range from the nanometer scale to a few thousand squared microns depending on the substrate used [38], [39]. Thus, the simulations presented here reflect the behavior of a typical sample size for a two-dimensional material. It is therefore physically reasonable to require that numerical delocalization cannot be claimed if the ratio $R$ contributes more than 10% of the computed distance parameter.

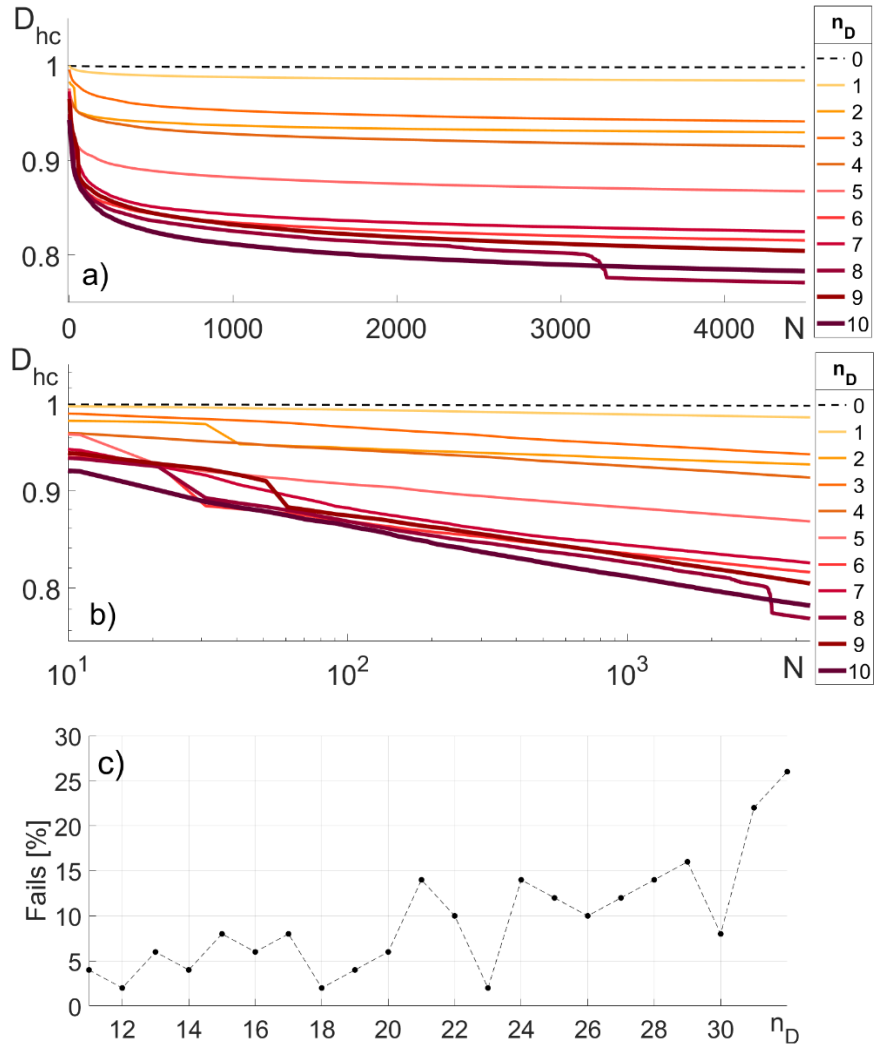

FIG. 7. Distance time evolution plots a) and the corresponding log-log plots b) for doping in the range $n_D(\%) = 1:1:10$. c) The number of realizations for which the distance dropped to zero for disorder in the range $n_D(\%) = 11:1:32$.

The values of $b$, $D(4500)$, and $R$ corresponding to doping concentrations $n_D(\%) = 1:1:10$ are provided in Table II. Due to the presence of failed realizations in this range, the calculation of the parameters $b$ and $R$ were obtained by first averaging all realizations per doping value, and then performing the equation fit [11]. In each case, the root mean squared error in the equation fitting analysis (showing the goodness of the fits) was $\lesssim 10^{-4}$. Comparison of columns two and three shows that, although the values of $D_{hc}$ at the last timestep are positive for all concentrations, the limiting values $b$ quickly drop and cross zero between $n_D(\%) = 5 - 6$. In addition, the contribution from the ratio $R$ for doping concentration $n_D(\%) = 1$ is already 22%, confirming that for the size of the performed simulations, $n_D(\%) \geq 1$ is already too large to yield distance parameters that flatten out exponentially. This further suggests that the 2D honeycomb lattice in this regime is not characterized by metallic behavior.

TABLE II. Extrapolated values from equation fitting of averaged distance plots corresponding to doping $n_D(\%) = 0{:}1{:}10$.

| $n_D(\%)$ | $b$ | $D(4500)$ | $R(\%)$ |
|---|---|---|---|
| 0 | 0.964 | 0.998 | 3% |
| 1 | 0.772 | 0.984 | 22% |
| 2 | 0.537 | 0.968 | 45% |
| 3 | 0.309 | 0.941 | 67% |
| 4 | 0.162 | 0.915 | 82% |
| 5 | 0.050 | 0.911 | 94% |
| 6 | -0.263 | 0.874 | 130% |
| 7 | -0.074 | 0.825 | 109% |
| 8 | -0.767 | 0.857 | 190% |
| 9 | -0.824 | 0.843 | 198% |
| 10 | -0.770 | 0.783 | 198% |

*2. Small variation of doping concentration*

The time evolution plots of the distance values obtained for the small doping range $n_D(\%) \in [0,1]$ are presented in Fig. 8a. For all concentrations, the averaged $D_{hc}$ was obtained using at least 30 realizations of the disorder, which minimizes possible numeric artifacts. As can be seen, the plots corresponding to $n_D(\%) = 0.9$ and $n_D(\%) = 1$ drop much faster than the rest, which suggests dissimilarity in their behavior. Analysis of the individual realizations used to obtain these two (averaged) distance time evolutions shows that in both cases, there is at least one realization that rapidly dropped to zero. Thus, based on the analysis of the previous section, it is reasonable to expect that the transition in the transport behavior of the lattice occurs in the even smaller range $n_D(\%) \in [0, 0.8]$. Figure 8b shows the log-log plots (solid lines) and the corresponding equation fits (dotted, dot-dashed, and dashed lines) obtained using equation (14). As expected, the slopes of the log-log plots become increasingly negative with increasing doping concentration, which indicates slower decay towards constant nonzero values.

---

[11] Alternatively, one can first perform a fit to each realization and then average the results, which will yield an error estimate (spread from the mean value) for $b$ and $R$. However, such analysis is only possible if there are no realizations that rapidly fall to zero, which is not the case for large doping in our study. In the small doping regime explored in the next section, the results are confirmed using both fitting techniques.

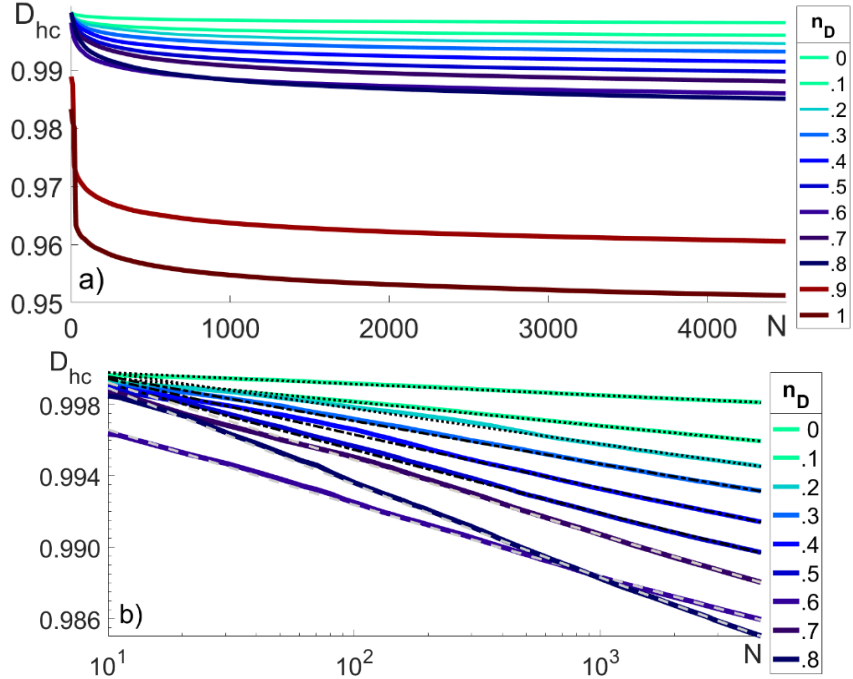

FIG. 8. Distance time evolution plots a) for the 2D honeycomb lattice with substitutional disorder in the range $n_D(\%) = 0:0.1:1$ and corresponding log-log plots b) for the range $n_D(\%) = 0:0.1:0.8$.

The averaged values for $b$, $D(4500)$, and $R$ are given in Table III. As mentioned above, concentrations $n_D(\%) = 0.9$ and $n_D(\%) = 1$ each had at least one realization that rapidly dropped to zero. Thus, the parameters for these cases were obtained by first averaging all of the realizations and then fitting to equation (14). In the smaller concentration range, $n_D(\%) \in [0,0.8]$, the values within parentheses reflect results from first performing the fits and then averaging, providing an error estimate based on the spread of the separate realizations. Although the two fitting techniques show agreement, the presence of outliers causes some differences in the values extrapolated for b and R. Since these outliers do not appear in a consistent manner as the concentration is increased (unlike the realizations that fall to zero for $n_D(\%) \geq 0.9$), it is not clear if they represent physically meaningful data or simply computational error. Thus, in this section we extrapolate $b$ and $R$ from the fits on the averaged distance values. A treatment that does not ignore outliers is presented in the next section, where the separate realizations are analyzed statistically.

For all cases considered, the root mean squared error [12] from the fit equation was $\sim 10^{-5}$, which indicates good agreement with the weighted equation fitting model. Table III shows that as doping concentration increases, the values of $b$ decrease, and the contribution $R$ becomes larger. For $n_D = 0.3$ (shaded row in Table III), the ratio $R$ contributes 10% of the distance value obtained at the last timestep, which suggests a transition in the behavior of the $D_{hc}$ plots.

---

[12] Note that there are two distinct errors in the discussion. The *error estimates* obtained from the spread of the random realizations for each disorder (the ones shown in Table II) indicate the certainty with which we can determine the limiting behavior of the distance values. The *root mean squared error* shows the goodness of the fit.

TABLE III. Extrapolated values from equation fitting of averaged distance plots corresponding to doping $n_D(\%) = 0: 0.1: 1$. The shaded row highlights the parameters for the critical value $n_D = 0.3\%$. The numbers in parentheses are the average and the standard deviation of each parameter based on the fits of individual realizations. As explained in the text, such analysis is only possible for the smaller range $n_D(\%) = 0: 0.1: 0.8$, where no $D$-values dropped to zero.

| $n_D(\%)$ | $b$ | $D(4500)$ | $R(\%)$ |
|---|---|---|---|
| 0 | 0.958 (960 ± 10) | 0.998 (998 ± 0) | 4 (4 ± 1) |
| 0.1 | 0.940 (942 ± 55) | 0.996 (996 ± 4) | 6 (5 ± 6) |
| 0.2 | 0.920 (920 ± 58) | 0.995 (995 ± 6) | 8 (7 ± 6) |
| 0.3 | 0.898 (881 ± 45) | 0.993 (993 ± 5) | 10 (9 ± 6) |
| 0.4 | 0.877 (860 ± 30) | 0.991 (991 ± 5) | 12 (11 ± 5) |
| 0.5 | 0.856 (850 ± 30) | 0.990 (930 ± 6) | 13 (13 ± 3) |
| 0.6 | 0.837 (827 ± 31) | 0.986 (986 ± 17) | 15 (14 ± 3) |
| 0.7 | 0.820 (796 ± 67) | 0.988 (988 ± 5) | 17 (16 ± 3) |
| 0.8 | 0.800 (796 ± 67) | 0.985 (985 ± 8) | 19 (19 ± 7) |
| 0.9 | 0.804 | 0.982 | 18 |
| 1 | 0.769 | 0.978 | 21 |

*3. Statistical analysis for small doping*

To confirm the existence of a transition point in the range $n_D(\%) = 0: 0.1: 0.8$, we study the dissimilarity among the values extrapolated for $b$ by equation fitting of the separate realizations for each concentration. The total number of observations in this range is 450 (i.e., 50 realizations per doping value). Figure 9 shows a dendrogram of the data obtained from hierarchical clustering algorithm with Ward's minimum variance method. Note that for clarity, the dendrogram consists of 30 leaves, where each leaf represents clusters of highly similar points. Fig. 9 shows the existence of two distinct clusters, which we name 'small group' and 'large group'. These clusters correspond to two different regimes in the limiting behavior of the distance plots, and the transition from one to the other is interpreted as a transition in the transport properties of the lattice.

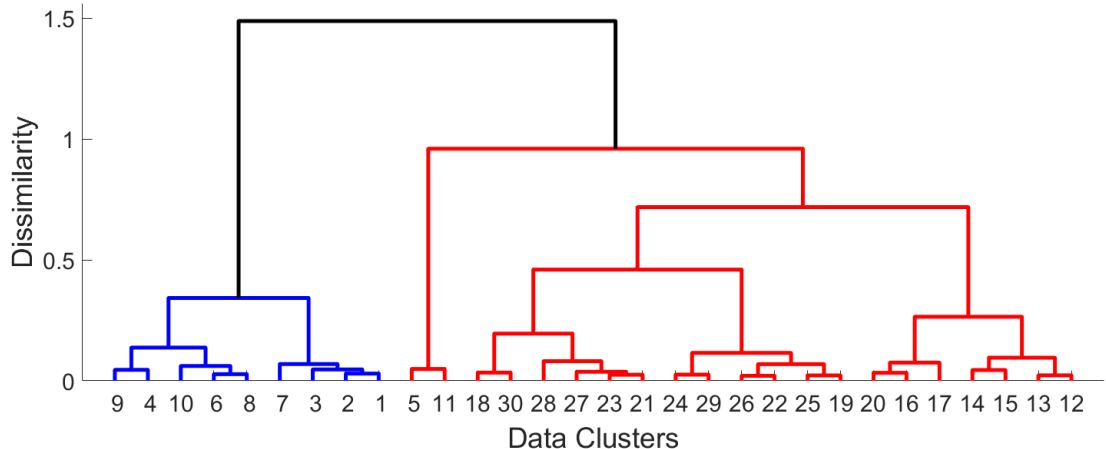

FIG. 9. Dendrogram for the averaged values of $b$ for doping in the range $n_D(\%) = 0:0.1:0.8$. Each of the 30 leaves represents clusters of highly similar data points. The blue cluster represents the small group, while the red cluster corresponds to the large group.

Table IV provides a classification of all examined $b$ values according to the two groups identified by clustering analysis. Comparison of the percent contributions to each group indicates that the transition occurs as the doping increases from $n_D = 0.3\%$ to $n_D = 0.4\%$, at which doping level a considerable percentage of realizations belongs to both groups. Indeed, a hypothesis test confirms that the average value of $b$ for doping levels from 0% to 0.3% inclusive is significantly different from the average value of $b$ for doping levels of 0.4% to 0.8% inclusive (p-value < 0.001). Thus, we conclude that $n_D = 0.3\%$ marks the onset of the insulating behavior in the lattice, in agreement with the experimental results presented by Bostwick et al. [8]. From Table IV we see that for doping values $n_D(\%) = 0.1, 0.2,$ and $0.6$, there is only one realization (or 2%) that lies in a different group, which is considered an outlier due to computational error.

TABLE IV. Percentage of observations for each doping level that are assigned to the "small" or "large" group by Hierarchical Clustering using Ward's Distance, as shown in Fig. 9.

| $n_D(\%)$ | 0 | 0.1 | 0.2 | 0.3 | 0.4 | 0.5 | 0.6 | 0.7 | 0.8 |
|---|---|---|---|---|---|---|---|---|---|
| Small Group | 100 | 98 | 98 | 78 | 20 | 0 | 2 | 0 | 0 |
| Large Group | 0 | 2 | 2 | 22 | 80 | 100 | 98 | 100 | 100 |

## IV. CONCLUSIONS

In this work, we have used the spectral approach (first introduced by Liaw in [13]) to establish the existence of extended states in a 2D honeycomb lattice with substitutional disorder. The examined transport problem is described by an Anderson-type Hamiltonian, where the hopping potential is represented by a constant nearest-neighbor interaction (tight-binding approximation), and the on-site energies are random variables assigned according to a predetermined probability distribution $\chi$. The hopping potential $V$ has been normalized to unity; thus, the energies discussed here are in units of $V$.

Doping has been introduced by assuming a modified bimodal distribution consisting of a Gaussian peak and a delta-type peak. The variables assigned from the Gaussian peak represent the fluctuations in the on-site energies of an unperturbed crystal. Physically, such fluctuations naturally occur due to finite temperature and spatial lattice defects. Preliminary analysis of various normal distributions suggest that it is reasonable to use a Gaussian peak with mean $E_A = 0$ and variance $\sigma^2 = 0.05$. In all numerical simulations, we used a delta function at $E_B \approx 100$ to represent the approximate energy of the doping atoms.

The defect concentration was varied between $n_D = 0\%$ and $n_D = 32\%$. For $n_D \geq 0.9\%$, the spectral approach does not predict the existence of extended states with probability 1. Thus, we conclude that in this range, the transport behavior of the system has transitioned into the localized regime. We applied an equation fitting model and hierarchical clustering for small concentrations of the doping, $n_D(\%) \in [0, 0.8]$. The results indicate the existence of a transition point in the transport behavior of the system for $0.3\% \leq n_D \leq 0.4\%$, which agrees with previous experimental results [8].

The most important limitation of this work is the finite character of the numerical simulations. The results from the present analysis can be improved by increasing the number of timesteps $N$, which will allow modeling of a larger crystal. This will yield better estimates for the errors in the values of $b$ and $R$, used to determine the existence of transitions in the transport behavior of the system. Further improvements of the present analysis will come from optimization of the parameters for the probability distribution $\chi$ so that specific application of 2D materials can be examined.

ACKNOWLEDGMENTS

This work was supported by the NSF-DMS (grant number 1802682, C D L), NASA grant number 1571701 and NSF grant numbers 1414523, 1740203, 1262031 (L S M and T W H).